\documentclass[conference]{IEEEtran}
\IEEEoverridecommandlockouts

\usepackage{amsmath,amssymb,amsfonts}
\usepackage{graphicx} \graphicspath{ {Figures/} }
\usepackage{tabularx}
\usepackage{url}
\usepackage{xcolor}
\usepackage{algorithm,algpseudocode}
\usepackage{nicefrac}
\usepackage{lipsum}

\begin{document}

%
\title{DeTRM: Decentralised Trust and Reputation Management for Blockchain-based Supply Chains
    \thanks{The first and second authors have the same contribution in the paper.}
}

\author{
    \IEEEauthorblockN{
    Guntur Dharma Putra\IEEEauthorrefmark{1}\IEEEauthorrefmark{3},
    Changhoon Kang\IEEEauthorrefmark{2},
    Salil S. Kanhere\IEEEauthorrefmark{1}\IEEEauthorrefmark{3}
    and James Won-Ki Hong\IEEEauthorrefmark{2}
}
\IEEEauthorblockA{
    \IEEEauthorrefmark{1}UNSW Sydney, Australia
    \IEEEauthorrefmark{2}POSTECH, South Korea
    \IEEEauthorrefmark{3}CSCRC, Australia
}
}

\maketitle


\begin{abstract}
Blockchain has the potential to enhance supply chain management systems by providing stronger assurance in transparency and traceability of traded commodities. However, blockchain does not overcome the inherent issues of data trust in IoT enabled supply chains. Recent proposals attempt to tackle these issues by incorporating generic trust and reputation management, which does not entirely address the complex challenges of supply chain operations and suffers from significant drawbacks. In this paper, we propose DeTRM, a decentralised trust and reputation management solution for supply chains, which considers complex supply chain operations, such as splitting or merging of product lots, to provide a coherent trust management solution. We resolve data trust by correlating empirical data from adjacent sensor nodes, using which the authenticity of data can be assessed. We design a consortium blockchain, where smart contracts play a significant role in quantifying trustworthiness as a numerical score from different perspectives. A proof-of-concept implementation in Hyperledger Fabric shows that DeTRM is feasible and only incurs relatively small overheads compared to the baseline.
\end{abstract}


\section{Introduction}
\label{sec:intro}
Supply Chain Management Systems (SCMS) aim to provide an end-to-end audit trail and provenance information of traded goods and services, which extends across the primary producers to end customers~\cite{juma2019}. In general, SCMS operate by creating a digital representation of physical commodities, referred to as digital assets. The subsequent transfers and ownership changes of the assets are recorded in the system, as they progress through the different stages of the supply chain~\cite{gonczol2020}. Internet of Things (IoT) devices have also been utilised in SCMS to streamline and automate supply chain processes by collecting relevant sensor readings from the environment, communicating with related parties and actuating certain actions as defined in the business logic~\cite{rejeb2019}. For instance, RFID tags have been widely adopted in warehousing and transportation of commodities to automate the tracking process~\cite{tajima2007}, while Wireless Sensor Networks (WSN) are often deployed to provide efficient monitoring of the commodities. 

Supply chains are experiencing exponential growth, which introduces unprecedented challenges in providing transparency and traceability for more complex participants, who may not necessarily trust each other~\cite{sahai2020}. Recently, blockchain has demonstrated its potential to enhance conventional SCMS with stronger assurance in traceability and transparency due to its inherent data structure of immutable time-stamped records~\cite{malik2019}. Blockchain also removes the need to trust a centralised entity to administer the SCMS, as trust is decentralised to all blockchain participants through a consensus mechanism. In addition, blockchain allows instant identification of the provenance of a commodity by searching through the linked blocks, which would take significantly longer time in conventional SCMS~\cite{gonczol2020}. Oftentimes this information is not available in conventional SCMS, due to the inherent data silos in centralised architectures~\cite{malik2018}. Additionally, smart contracts can also be utilised to record trading agreements between supply chain participants in a secure and verifiable manner.

However, relying solely on blockchain for IoT supported SCMS does not solve the underlying trust problem associated with the data. In fact, sensor nodes may become faulty and send inaccurate and low-quality data, which would become immutable once stored on the blockchain, contaminating the ledger with bad data~\cite{powell2021a}. Trust and Reputation Management (TRM) is an effective solution to overcome this issue. However, existing approaches for incorporating TRM into blockchain-based SCMS, e.g.,~\cite{malik2019, shahid2020, li2020a} adopt a generic TRM model, which does not cater to the unique challenges of SCMS, such as dynamic operations in production of commodities. In addition, the existing work assumes that the commodities always move across the supply chain in a fixed way, which is unlikely as supply chains predominantly involve productions of new commodities from different sources~\cite{zhang2021}.

In this paper, we propose DeTRM, a decentralised TRM framework for blockchain-based SCMS, which aims to overcome trust issues in data and behaviour of supply chain participants. DeTRM quantifies the quality of sensor observations to provide a numerical measure of data trust, while the behaviour of supply chain participants is constantly evaluated as per pre-determined trade agreements to assess their trustworthiness. We define distinct scores for trust and reputation, using which the consumers can conveniently infer trust from different levels of perspective. In addition, we incorporate inherent supply chain operations, e.g., production of commodities, in designing our tailored TRM model for SCMS to keep track of trust information across the entire supply chain life cycle, i.e., from sourcing the raw material to the retail shelf. In DeTRM, each supply chain entity is required to host blockchain peer nodes to build and maintain a consortium blockchain, enabling a firm decentralised framework. We implemented a proof-of-concept implementation of DeTRM in a lab-scale Hyperledger Fabric network, and evaluated the evolution of trust in typical usage scenarios and benchmarked the incurred overheads. Experimental results show that DeTRM incurs minimal overheads compared to the baseline with regard to resource utilisation, throughput and latency.

In summary, we make the following contributions in this paper:
\begin{itemize}
    \item We propose DeTRM, a decentralised TRM framework for blockchain-based SCMS, in which we consider complex supply chain operations such as splitting or merging of product lots or processing raw materials to produce a new product while modelling the TRM.
    \item DeTRM resolves data trust by correlating observational data from adjacent sensor nodes, using which the integrity and authenticity of the data can be assessed and corresponding scores can be quantified.
    \item DeTRM provides distinction of trust and reputation scores, which offers a convenient way to determine the trustworthiness of supply chain participants from different perspectives. Here, smart contracts are utilised to enable transparent and verifiable mechanisms to compute these scores. 
    \item We develop a proof-of-concept implementation on Hyperledger Fabric network hosted on a lab-scale testbed. We benchmark DeTRM to demonstrate its effectiveness in tracking the evolution of trust evolution and determine performance metrics such as transaction throughput and latency. Experimental results indicate that DeTRM is feasible and only incurs minimal overheads compared to the baseline.
\end{itemize}

The rest of the paper is organised as follows. Section~\ref{sec:related-work} discusses the related work, while Section~\ref{sec:system-model} describes the proposed system model. We elaborate on the proposed TRM in Section~\ref{sec:trust-reputation-systems}, while the decentralised supply chain framework is elaborated in Section~\ref{sec:supply-chain-framework}. We present the performance evaluation in Section~\ref{sec:performance-evaluation}. We conclude this paper and present future work in Section~\ref{sec:conclusion}.

\section{Related Work}
\label{sec:related-work}
Blockchain has received a significant amount of attention from the research community as a fundamental building block to build secure and trustworthy SCMS, due to its inherent characteristics, such as immutability and auditability. In~\cite{malik2018}, the authors proposed a generic framework for blockchain-based Food Supply Chain (FSC), in which a consortium blockchain acts as a platform for FSC participants to provide end-to-end provenance of the traded products. The framework is comprised of a tiered architecture with a built-in support to distribute the blockchain into multiple shards for maintaining scalability. Zhang et al. proposed a blockchain-based supply chain framework with an emphasis on process adaptation~\cite{zhang2021}, which allows flexibility in defining complex supply chain process as a collection of fragments.  In~\cite{sahai2020}, the authors proposed a privacy-preserving framework for achieving provenance in SCMS, where contamination tracing is introduced to identify contaminated products in the supply chain. However, these works~\cite{malik2018,zhang2021,sahai2020} disregard the importance of incorporating trust management to determine the authenticity of the incoming data and to evaluate the reliability of supply chain participants.

Researchers have proposed the notion of blockchain-based TRM, where blockchain is incorporated to provide trustless and secure evaluation of participants behaviour and the quality of the exchanged data~\cite{putra2021a}. For instance, TRM is coupled with a decentralised Attribute-based Access Control (ABAC) to quantitatively assess the trustworthiness of network participants and help protect the network from adversaries~\cite{putra2021b}. TRM is also implemented in an IoT data trading scenario to provide assurance in the reliability of sellers by means of numerical measures, using which the consumers can select appropriate sellers based on their previous behaviour~\cite{camilo2020}. Unfortunately, these decentralised TRM systems~\cite{putra2021a, camilo2020} cannot be implemented directly to SCMS, due to inherent characteristics of supply chain that requires complex operations of the traded goods and commodities, such as splitting and repackaging of raw materials.

Some effort to incorporate TRM into SCMS have been proposed in~\cite{malik2019, shahid2020, li2020a}. A rudimentary TRM for supply chain in agriculture is proposed in~\cite{shahid2020}, while a smart contract-based reputation model for SCMS is proposed by Li et al.~\cite{li2020a}, in which tokens are used as the proxy of reputation. However, these approaches~\cite{shahid2020, li2020a} incorporate streamlined TRM concepts and disregard complex operations of supply chain scenarios. TrustChain~\cite{malik2019} is by far one of the few frameworks that tailors a TRM specifically for the case of decentralised SCMS, where the reputation of a supply chain participant is determined from multiple sources. However, the framework still suffers from several shortcomings. First, TrustChain fully delegates the administrative control and management of the blockchain network to the business network administrator, essentially a Trusted Third Party (TTP), which undermines the fundamental motivation of using a blockchain. Second, the framework's TRM model is directly adopted from a generic model~\cite{moinet2017}, which may not accurately resemble real-world supply chain scenarios. Lastly, TrustChain relies on a strong assumption that the commodities are static and will not be reproduced or repackaged as they progress through the supply chain life cycle. However, in practice commodities would very likely be split and repackaged through the supply chain life cycle.

In summary, previous work in the literature fail to provide a comprehensive TRM solution for blockchain-based SCMS. In addition, the trust model is either oversimplified or directly adopted from generic trust model without proper modifications for complex operations in supply chain systems. DeTRM is designed to overcome the aforementioned shortcomings and to provide a coherent solution for trust management in blockchain-based SCMS.

\begin{figure}
\centering
\includegraphics[width=0.47\textwidth]{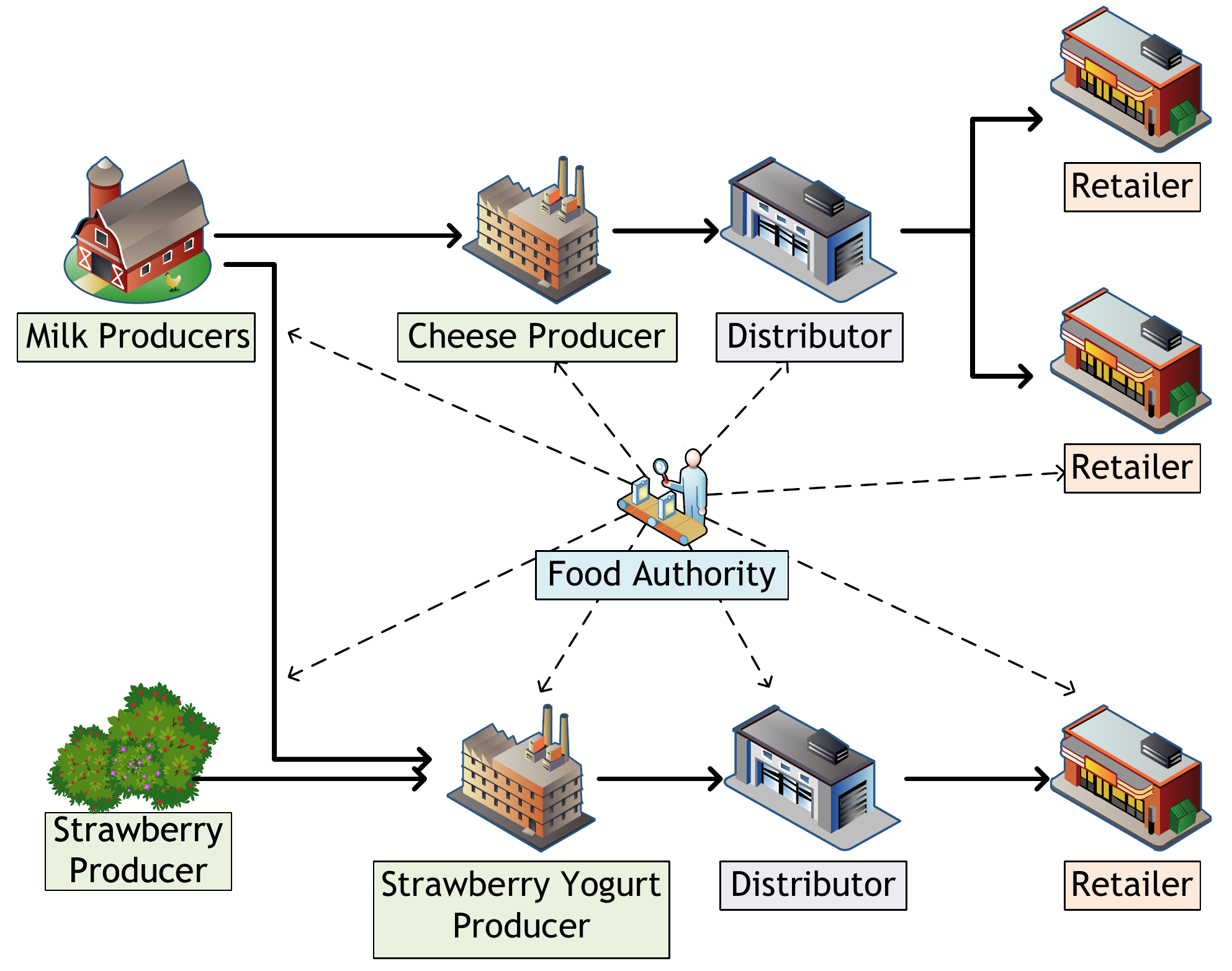}
\caption{An overview of a simplified dairy supply chain from primary producers to retail stores. The food authority here is regulating the supply chain and ensuring that the commodities are up to the safety standard.}
\label{fig:dairy-overview}
\end{figure}

\section{System Model}
\label{sec:system-model}
In this paper, we model our SCMS using diary supply chain as a representative example.
However, our model can also be generalised and applied to other supply chain use cases. Fig.~\ref{fig:dairy-overview} depicts an example of diary supply chain, which generally consists of producers, distributors, retailers and a food authority.

\subsection{Supply Chain Entities}
\label{sub:supply-chain-entities}
Our decentralised SCMS model, denoted $\mathbb{SC}$, defines two types of supply chain entities, denoted $\mathbb{E} = (A, P)$, where $A=\{a_1,\ldots,a_m\}$ is a set of regulatory authorities and $P=\{p_1,\ldots,p_n\}$ is a set of supply chain participants. In our model, a participant $p_n \in P$ is represented as a tuple $p_n = \left<\mathcal{P}^n_{id}, \mathcal{P}^n_{role}, \left<\mathcal{P}^n_{prop}\right>\right>$, where $\mathcal{P}^n_{id}$ is the unique ID of participant $p_n$, $\mathcal{P}^n_{role}$, is the participant's role in the supply chain and  $\left<\mathcal{P}^n_{prop}\right>$ is a set of properties for $p_n$. We define three types of role for participant $p_n$, namely \textit{producers}, \textit{distributors} and \textit{retailers}, denoted $\mathcal{P}^n_{role} \subseteq \{\mathit{producer}, \mathit{distributor}, \mathit{retailer}\}$. As such, a participant $p_n \in P$ can uniquely be a producer or can have multiple roles at the same time, e.g., be both a distributor and retailer. We exclude the consumers from the set of participants $P$, as consumers are not required to participate and maintain blockchain network $\mathbb{B}$. However, they are allowed to access $\mathbb{B}$, for example, to see the trail of supply chain and its trust scores by using an imprinted QR code on the product. We present a table of notations in Table~\ref{ta:notation-table}.

    \textbf{Food Authorities:}
    The food authorities, such as NSW Food Authority\footnote{\url{https://www.foodauthority.nsw.gov.au/}}, are mainly responsible to periodically perform regulatory inspections regarding the safety of the diary products, according to a particular safety and regulatory standard, e.g., HACCP~\cite{mortimore2013haccp}.
    
    \textbf{Supply Chain Participants:}
    In our model, the participants manage the creation and transfer of digital assets that represent real physical commodities being traded in the SCMS. In this paper, we refer to commodities and digital assets interchangeably, as they refer to the same object. A digital asset owned by participant $p_n$ is denoted as $d_{n,q} = \left<\mathcal{D}^{n,q}_{owner}, \left<\mathcal{D}^{n,q}_{prop}\right>\right>$, where $\mathcal{D}^{n,q}_{owner}$ is the ID of the current owner and $\mathcal{D}^{n,q}_{prop}$ is a set of properties for $d_{n,q}$, including the batch ID. We denote a set of digital assets owned by $p_n$ as $D_n=\{d_{n,1}, d_{n,2}, \ldots, d_{n,q}\}$. Specifically, \textit{producers} are entities that make the products or commodities, sourced from raw materials or from existing commodities in $\mathbb{SC}$, while \textit{distributors} purchase several commodities from producers in large quantities for resale purpose. We group shipping and logistic entities, who are in charge for product transportation into this category. On the other hand, \textit{retailers} are the entities that trade commodities to individual consumers for consumption in a relatively smaller quantity.
    
\renewcommand{\arraystretch}{1.2}
\begin{table}
\centering
\caption{Summary of important notations and their description}
\label{ta:notation-table}
\begin{tabular}{|c|l|}
\hline
\textit{\textbf{Notations}} & \textit{\textbf{Description}} \\
\hline \hline
$\mathbb{SC}$; $\mathbb{B}$ & supply chain framework; blockchain network \\
\hline
$\mathbb{E} = (A, P)$ & supply chain entity: authority and participant \\
\hline
$a_m \in A$ & food authority $m$ \\
\hline
$p_n \in A$ & supply chain participant $n$ \\
\hline
$d_{n,q} \in D_n$ & digital asset $a$ owned by $p_n$ \\
\hline
$\ell_k \in L_n$ & locations of participant's premises \\
\hline
$s_{n,p} \in S_n$ & IoT sensor node $p$ \\
\hline
$\textsl{g}_{n,k}$ & gateway node at location $\ell_k$ \\
\hline
$v^n_{o,p} \in \mathbf{v}^{n}_o$ & measurement data \\
\hline
$\mathcal{V}_{o,p}$, $\mathcal{V}^C_{o,p}$ & the value and confidence of an observation \\
\hline
$e_{m,q}$ & endorsement rating from $a_m$ \\
\hline
$\mathcal{C}_{trm}$ \& $\mathcal{C}_{com}$ & TRM and commodity smart contracts \\
\hline
$\widehat{t}_{n,q}$ & trust score of a commodity $d_{n,q}$ \\
\hline
$\widehat{T}_n$ & trust score of a participant $p_n$ \\
\hline
$\widehat{R}_n$ & reputation of a participant $p_n$ \\
\hline
$\gamma$ & decaying parameter for trust and reputation \\
\hline
\end{tabular}
\end{table}

\subsection{Layered View of the System}
\label{sub:layer-view}
\begin{figure}
\centering
\includegraphics[width=0.5\textwidth]{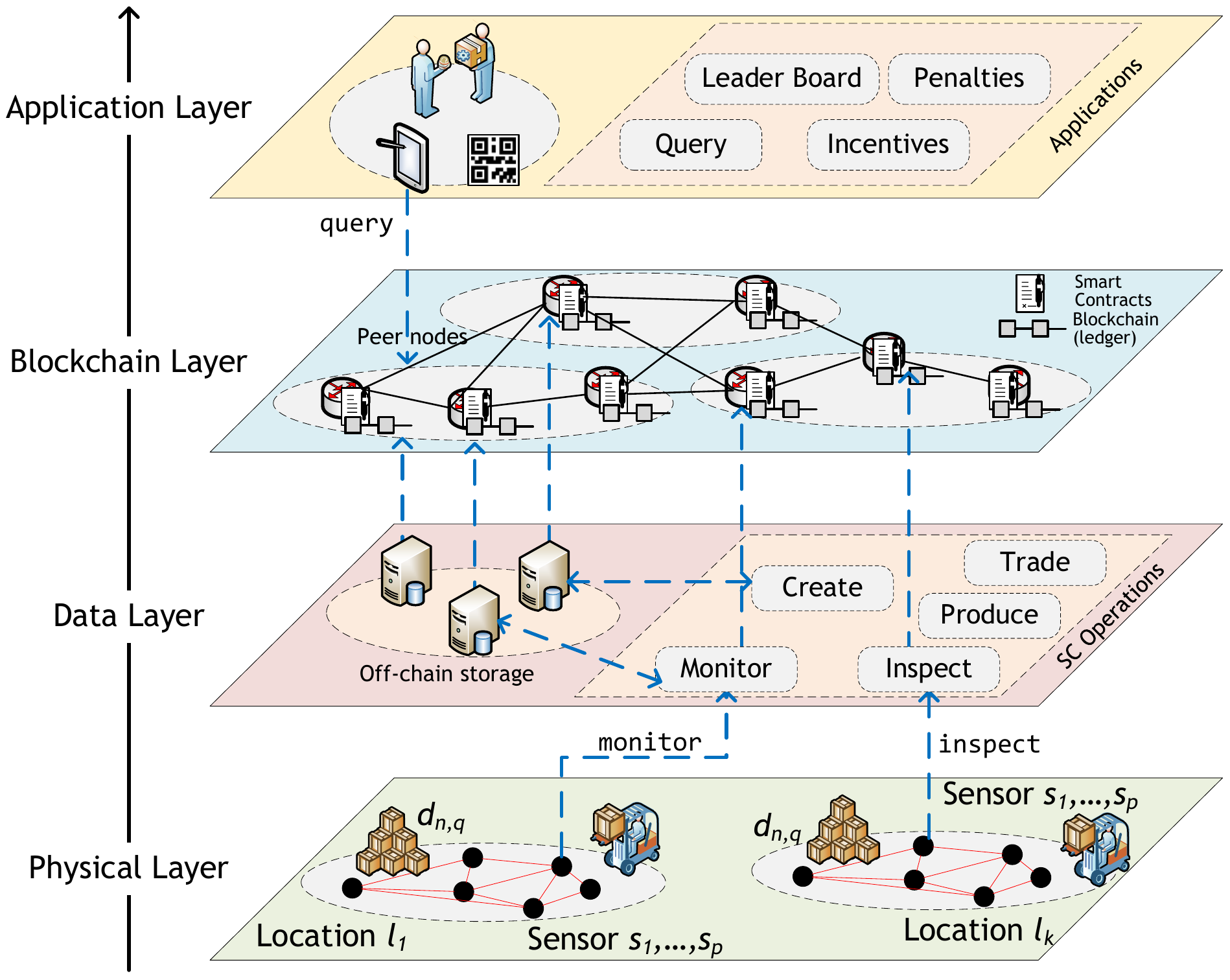}
\caption{The layered overview of the proposed system model, which consists of four interconnected layers namely physical, data, blockchain and application layer.}
\label{fig:system-model}
\end{figure}

Fig.~\ref{fig:system-model} presents a layered view of DeTRM. We categorise the main components in $\mathbb{SC}$ into four interconnected layers, denoted $\mathbb{SC}=\left<\mathbb{L}_{app}, \mathbb{L}_{bc}, \mathbb{L}_{data}, \mathbb{L}_{phy}\right>$, namely application, blockchain, data and physical layer.

    \textbf{Physical layer $\mathbb{L}_{phy}$:}
    In our model, a participant $p_n$ may store their assets $D_n$ in their premises, denoted $L_n=\{\ell_1,\ldots,\ell_k\}$ where $\ell_k=\left<\mathcal{L}_{lat}, \mathcal{L}_{lon}\right>$ corresponds to the location of the premises. To monitor the assets in each $\ell_k$, participant $p_n$ deploys several IoT sensor nodes, denoted $S_n=\{s_{n,1}, \ldots, s_{n,p}\}$. A sensor node $s_{n,p} \in S_n$ is represented by a tuple $s_{n,p} = \left<\mathcal{S}_{key}, \mathcal{S}_{mod}, \mathcal{S}_{owner}\right>$, where $\mathcal{S}_{key}$ is a unique sensor ID, $\mathcal{S}_{mod}$ refers to sensor modality and $\mathcal{S}_{owner}$ is the ID of $p_n$. Function $f_{\mathcal{M}} : S_n \rightarrow L_n$ maps each sensor node $s_{n,p} \in S_n$ to a particular location $\ell_k \in L_n$, resulting in a unique set $\mathcal{M}_k \subseteq S_n$ that represents a set of IoT sensor nodes in location $\ell_k$. To guarantee data authenticity and avoid erroneous sensory data, we require that each location $\ell_k$ is supplied with at least $y$ sensor nodes with similar $\mathcal{S}_{mod}$ providing sensor redundancy. We assume a set of sensor nodes $\mathcal{M}_k$ is connected to a gateway $\textsl{g}_{n,k}$, which has enough resources for performing asymmetric cryptography to communicate with blockchain $\mathbb{B}$. In our model, we assume that the participants use Time Temperature Indicator (TTI) sensors. However, our model can also be generalised to other types of sensor such as GPS and humidity sensor, depending on the application scenario.
    
    \textbf{Data layer $\mathbb{L}_{data}$:}
    Supply chain participants $P$ perform a set of operations for managing the creation, movements and transfer of assets, which are recorded via IoT sensor nodes $\mathbf{s}_{n,p}$, which then stored on the blockchain $\mathbb{B}$. We define five types of operations based on the following supply chain events:
    
    \begin{enumerate}
        \item \textit{Create:}
        Producers $\{ p_n \in P \mid P^x_{role} = \mathit{producer}\}$ are required to create digital assets for digitising physical commodities in $\mathbb{SC}$. A $\mathsf{create}$ operation results in the creation of a new digital asset $d_{n,q}$ from producer $p_n$, which is formally defined as follows:
        \begin{equation}
        \label{eq:data-create}
            d_{n,q} \leftarrow \mathsf{create}(\mathcal{D}^{n,q}_{\mathit{owner}}, \left<\mathcal{D}^{n,q}_{\mathit{prop}}\right>) \text{.}
        \end{equation}
        For instance, a milk producer creates a new digital asset for each batch of raw milk, where the properties of each raw milk batch are manifested in $\left<\mathcal{D}^{n,q}_{\mathit{prop}}\right>$.
        
        \item \textit{Produce:}
        To record new commodity production in $\mathbb{SC}$, producer $p_n$ is required to perform $\mathsf{produce}$ operation which is defined as:
        \begin{equation}
        \label{eq:data-produce}
            d^{\prime}_{n,q} \leftarrow \mathsf{produce}(D_{\mathbb{S}}, \mathbf{R}^p)
        \end{equation}
        where $d^{\prime}_{n,q}$ is the new digital asset, e.g., strawberry yogurt, $D_\mathbb{S}=\{d_{n,q}, d_{n,(q+1)}\}$ is a set of raw materials and $\mathbf{R}^p$ is a set of required parameters the process. Note that $\mathsf{produce}$ creates new digital assets from existing ones, while $\mathsf{create}$ imports new unrecorded commodities into $\mathbb{SC}$ as new digital assets.
        
        \item \textit{Monitor:}
        IoT sensor nodes perform $\mathsf{monitor}$ operations on a regular basis to check if the commodity is stored within acceptable temperature range as per the safety standard. A sensor node $s_{n,p}$ submits measurement data for observation $o$, denoted as a tuple $v^n_{o,p}=\left<\mathcal{V}_{o,p}, \mathcal{V}^C_{o,p}\right>$, where $\mathcal{V}_o$ refers to measurement value, e.g., temperature scale ($^{\circ}C$), and $\mathcal{V}^c_o\in [0,1]$ refers to the confidence of the observation, using which the quality of observational data can be derived~\cite{frolik2001}. We formally define $\mathsf{monitor}$ operation as:
        \begin{equation}
        \label{eq:data-monitor}
            d_{n,q} \leftarrow \mathsf{monitor}(\textsl{g}_{n,k},\mathbf{v}^{n}_o,\ell_k)
        \end{equation}
        where $\mathbf{v}^{n}_o=\{v^n_{o,1},v^n_{o,2},\ldots,v^n_{o,p}\}$ is a set of measurement data of observation $o$ from $p$ sensor nodes in location $\ell_k$. Here, $\mathsf{monitor}$ operation is performed by gateway $\textsl{g}_{n,k}$, to which each sensor node $s_n \in \mathcal{M}_k$ submits the measurement value.
        
        \item \textit{Inspect:}
        Food authorities are required to regularly conduct on-site checks of the production, storage and transportation facilities to ensure the quality of the products. On completion, food authority $a_m$ records the results to $\mathbb{SC}$ via $\mathsf{inspect}$ operation:
        \begin{equation}
        \label{eq:data-inspect}
            \left<d_{n,q}, p_n\right> \leftarrow \mathsf{inspect}(a_m,e_{m,q},\mathbf{R}^q)
        \end{equation}
        where $e_{m,q} \in [0,1]$ is the endorsement rating from $a_m$ according to inspection results and $\mathbf{R}^q$ is a set of detailed parameters pertaining to the on-site checks.
        
        \item \textit{Trade:}
        This operation corresponds to a change of asset ownership between two supply chain participants. For each $\mathsf{trade}$ operation, both participants should agree on a set of terms and conditions (trade agreements), denoted $\mathbf{tc}_x$, such as payment and shipment due date. A $\mathsf{trade}$ operation over an exchange of asset $d_{n,q}$ from participant $p_n$ to $p_{(n+1)}$ is defined as follows:
        \begin{equation}
        \label{eq:data-trade}
            d^{\prime}_{(n+1),q} \leftarrow \mathsf{trade}(d_{n,q}, p_n, p_{(n+1)}, \mathbf{tc}_x)
        \end{equation}
        where $d^{\prime}_{(n+1),q}$ denotes a new digital asset owned by participant $p_{(n+1)}$.
        
    \end{enumerate}
    
    Note that these operations are coupled with complementing blockchain transactions (see Section~\ref{sec:supply-chain-framework}). In addition, layer $\mathbb{L}_{data}$ also holds meta-data for supporting data collection (e.g., inspection paperwork and certification), which can be hosted on a centralised or peer-to-peer storage, as long as they are accessible by supply chain participants. We require the hash of the corresponding data to be stored on chain for maintaining integrity.

    \textbf{Blockchain layer $\mathbb{L}_{bc}$:}
    We adopt a modular permissioned-blockchain design~\cite{androulaki2018}, where supply chain entities $\mathbb{E}$ collectively establish and maintain a consortium blockchain network $\mathbb{B}$. Each food authority $a \in A$ and supply chain participant $p \in P$ host blockchain peer nodes, through which they interact with other entities. Each peer node is identifiable by a pair of private (secret) and public key, denoted $k_s, k_p$. We consider a modular design of smart contracts, where we design two smart contracts, namely TRM contract, denoted $\mathcal{C}_{trm}$, and commodity contract, denoted $\mathcal{C}_{com}$. Trade agreements and the trust model are stored in $\mathcal{C}_{trm}$, while temperature threshold, time interval $\Delta\tau$ and other quality related information are stored in $\mathcal{C}_{com}$. Note that our model is blockchain agnostic and can be implemented on any permissioned blockchain platform that supports smart contracts execution, for instance Hyperledger Fabric~\cite{androulaki2018} and Corda~\cite{brown2016corda}.

    \textbf{Application layer $\mathbb{L}_{app}$:}
    This layer mainly provides an interface for customers and other relevant parties to benefit from the TRM, e.g., querying the scores for each seller or commodities. In this layer, Food Authorities $A$ may host a REST API, through which customers can query for the reputation of each supply chain participant, e.g., by scanning a QR code displayed on the traded products.

In our model, we assume that food authority is trusted, while supply chain participants are not necessarily trusted.
We further assume that DeTRM inherits the assumptions of a commodity blockchain platform, which include security against peer-to-peer and consensus attacks, such as Sybil, eclipse and 51\% attacks~\cite{8543246}. All peer nodes in $\mathbb{SC}$ are bound to cryptographic primitives, which prevents manipulation and duplication of blockchain identities, i.e., public and private key pairs.

\begin{figure*}[!t]
\centering
\includegraphics[width=0.9\textwidth]{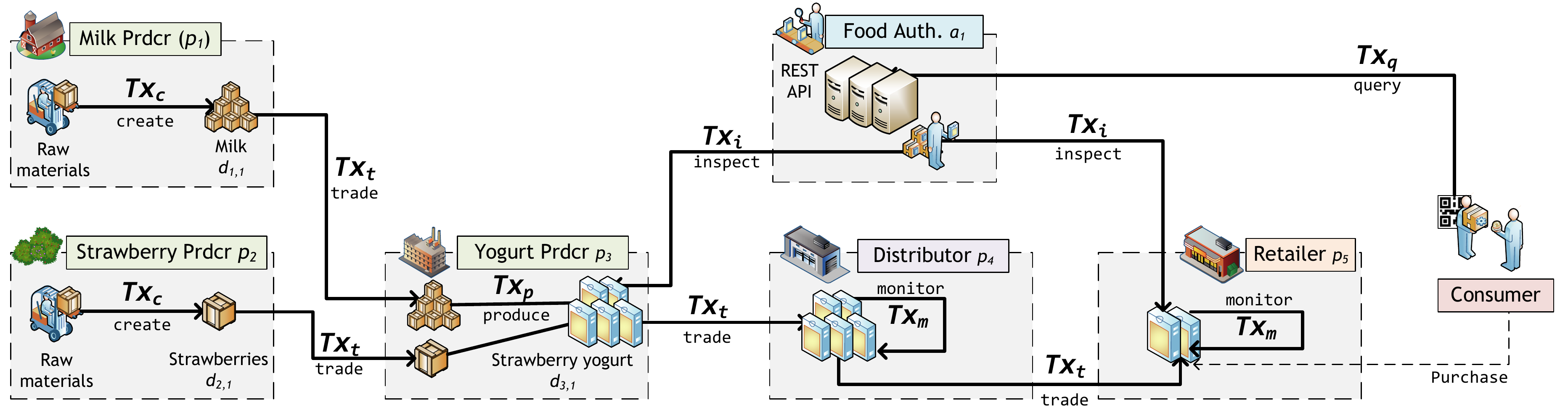}
\caption{An example of the workflow of DeTRM, beginning from product creation, trade, production, monitoring, inspection and query. The supply chain entities $\mathbb{E} = \{a \in A, p \in P\}$ are shown in a dashed rectangle. The consumers are not considered as supply chain entities.}
\label{fig:sequence-diagram}
\end{figure*}

\section{Trust and Reputation Management Model}
\label{sec:trust-reputation-systems}
The main goal of DeTRM is to evaluate the trustworthiness of the supply chain participants and to provide quality assurance of the traded commodities. We assign trust and reputation scores for each commodity $d_{n,q}$ and supply chain participant $p_n$ in $\mathbb{SC}$ to provide a quantified measure of trustworthiness. The scores are mainly derived from the operations in data layer $\mathbb{L}_{data}$, which includes three major supply chain events, namely $\mathsf{monitor}$, $\mathsf{inspect}$ and $\mathsf{trade}$.


In general, the trust score of a commodity $d_{n,q}$ is derived from $\mathsf{monitor}$ operations (i.e., entity-based trust). The trust score of $d_{n,q}$, denoted $\widehat{t}_{n,q}$, is gradually increased when $\mathsf{monitor}$ operation reports expected temperature readings, i.e., $\mathrm{T}_{min} < \mathcal{V}_{o,p} < \mathrm{T}_{max}$, while $\widehat{t}_{n,q}$ is significantly decreased when $\mathsf{monitor}$ operation reports otherwise. The value of $\widehat{t}_{n,q}$ is derived from a set of $o$ observations from $p$ sensor nodes $\{s_n \mid \forall s_n \in \mathcal{M}_k\}$ at location $\ell_k$ arranged in an $o \times p$ matrix:
\begin{equation*}
\mathbf{V}^n = \begin{bmatrix} 
    v^n_{11} & v^n_{12} & \dots  & v^n_{1p} \\
    v^n_{21} & v^n_{22} & \dots  & v^n_{2p} \\
    \vdots   & \vdots   & \ddots & \vdots \\
    v^n_{o1} & v^n_{o2} & \dots  & v^n_{op} \\
    \end{bmatrix}
\end{equation*}
and $\widehat{t}_{n,q}$ is calculated as follows:
\begin{equation}
\label{eq:trust-for-assets}
    \widehat{t}_{n,q}(o,p) = \frac{(1 - \gamma)}{p} \sum^{o}_{i=1}\sum^{p}_{j=1}\gamma^{(o - i)}\delta_{j,i} \mathcal{V}^C_{j,i} \mathcal{V}^E_{j,i}
\end{equation}
where
\begin{equation*}
    \delta_{j,i} = \begin{cases} 
            \delta_{max}, & \text{if\ } \mathrm{T}_{min} < \mathcal{V}_{j,i} < \mathrm{T}_{max} \text{\ ,} \\
            \delta_{min}, & \text{otherwise,}
        \end{cases}
\end{equation*}
where $\delta_{max}$ and $\delta_{min}$ are the weights associated with temperature reading within and outside expected threshold, respectively. In addition, $\delta_{max}$ serves as the upper limit of $\widehat{t}_{n,q}$, while $\delta_{min}$ sets the lower boundary, i.e., $\widehat{t}_{n,q} \in [\delta_{min},\delta_{max}]$. We introduce $\gamma$ as a decaying constant ($0<\gamma\leq1$) to afford higher weights to recent observations. $\mathcal{V}^E_{j,i} \in [0,1]$ is the measurement evidence~\cite{dedeoglu2019} determined via correlation with other neighbouring sensor nodes $\mathcal{M}^\prime_k=\{s_{n,p} \in \mathcal{M}_k \mid s_{n,p} \neq s_{n,j}\}$:
\begin{equation}
\label{eq:trust-evidence}
    \mathcal{V}^E_{j,i} = \frac{1}{| \mathcal{M}^\prime_k |} \sum_{s_{n,p} \in \mathcal{M}^\prime_k}  \varphi_{n,p} \mathcal{V}^C_{o,p}
\end{equation}
where
\begin{equation*}
    \varphi_{n,p} = \begin{cases} 
            \phantom{-}1, & \text{if\ } \mathcal{V}_{o,p} \text{\ supports\ } \mathcal{V}_{j,i} \\
                      -1, & \text{otherwise}
        \end{cases}
\end{equation*}
where $|\mathcal{M}^\prime_k|$ denotes the cardinality of set $\mathcal{M}^\prime_k$, i.e., the number of neighbouring sensor nodes. Intuitively, the value of $\widehat{t}_{n,q}$ is directly proportional to the observation confidence $\mathcal{V}^C_{j,i}$ and evidence $\mathcal{V}^E_{j,i}$, in which an observation with high confidence score, supported with high evidence value would increase $\widehat{t}_{n,q}$ and vice versa. Consequently, participant $p_n$ would have a set of $q$ trust scores, denoted $\widehat{\mathbf{t}}_n = \{\widehat{t}_{n,1},\ldots,\widehat{t}_{n,q}\}$, for all of their commodities.

The trust score for participant $p_n$ (i.e., behaviour-based trust) is built gradually from each $\mathsf{trade}$ operation. $\mathcal{C}_{trm}$ evaluates how $p_n$ fulfils trade agreements in $\mathbf{tc}_x$ by quantifying it into a scalar value $\sigma_i \in [0,1]$. The score trust score $\widehat{T}_n$ after $r$ $\mathsf{trade}$ operations is calculated as follows:
\begin{equation}
\label{eq:participant-trust}
    \widehat{T}_n(r) = (1-\gamma)\sum^{r}_{i=1}\gamma^{(r-i)}\sigma_i
\end{equation}
where
\begin{equation*}
    \sigma_i = \frac{1}{|\mathbf{tc}_i|} \sum_{tc_j \in \mathbf{tc}_i} \psi_{tc_j}  tc_j
\end{equation*}
where $tc_j \in [0,1]$ is the score for agreement $j$ and $\psi_{tc_j} \in \{1, -1\}$ is the weight to indicate a successful ($1$) or unsuccessful ($-1$) fulfilment, which is automated by smart contracts $\mathcal{C}_{trm}$. In general, the more successful $\mathsf{trade}$ operations participant $p_n$ has, the higher trust score $\widehat{T}_n$ grows.

The reputation score of a participant $p_n$, denoted $\widehat{R}_n$, is a weighted average of entity and behaviour-based trust scores, cf.~\eqref{eq:trust-for-assets},~\eqref{eq:participant-trust}, and aggregated endorsement scores from food authorities. We calculate the score $\widehat{R}_n$ for $q$ commodities, $r$ $\mathsf{trade}$ operations and $u$ endorsements as follows:
\begin{equation}
\label{eq:participant-reputation}
    \widehat{R}_n(q, r, u) = \frac{w_t}{|\widehat{\mathbf{t}}_n|}\sum_{\widehat{t}_{n,q} \in \widehat{\mathbf{t}}_n}\widehat{t}_{n,q} + w_T \widehat{T}_n(r) +  w_e \widehat{E}_n(u)
\end{equation}
where
\begin{equation}
\label{eq:endorsement-formula}
    \widehat{E}_n(u) = (1-\gamma)\sum^{u}_{i=1}\gamma^{(u-i)}e_i
\end{equation}
where $e_i$ is the endorsement rating, and $w_t$, $w_T$ and $w_e$ refer to the weight for different reputation components (i.e., $\widehat{t}_{n,q}$, $\widehat{T}_n$ and $\widehat{E}_n$) and we require that $w_t + w_T + w_e = 1$. How we derive the weight depends on the setting.

All trust computation models are manifested in the $\mathcal{C}_{trm}$ smart contract to deliver an automated and secure decentralised TRM. In addition, the corresponding evidence and data source are logged in $\mathcal{C}_{trm}$, which ensures transparency and immutable audit trail of trust and reputation score evolution. Subsequently, supply chain participants and other related parties can conveniently query the contract $\mathcal{C}_{trm}$ to get the latest score of a particular commodity or participant.

\section{TRM Framework for Supply Chain}
\label{sec:supply-chain-framework}
In this section, we describe DeTRM with regard to supply chain operations in our system model and their blockchain transactions counterpart, cf. Section~\ref{sec:system-model}. In addition, this section presents how the trust calculation model is implemented in practice. Please refer to Table~\ref{ta:notation-table} for a summary of notations and their descriptions.

To initialise $\mathbb{SC}$, food authorities $A$ are entrusted to configure and initiate a consortium blockchain network $\mathbb{B}$, including policy rules for network governance, e.g, allowing and revoking network access. Food authorities are also entrusted to deploy $\mathcal{C}_{trm}$ contract as the base smart contract in $\mathbb{SC}$. Subsequently, supply chain participant $s_n \in S$ can join the network by sending a join request, denoted $Req_n = \left<\mathcal{P}^n_{id}, \mathcal{P}^n_{role}, \left<\mathcal{P}^n_{prop}\right>\right>$, to a food authority $a_m$ over a secure channel~\cite{dierks2008transport}. Upon successful validation, food authority $a_m$ provides $s_n$ with corresponding network configurations to join blockchain $\mathbb{B}$. To participate in the network, each supply chain entity is required to run blockchain peer nodes with a pair of public and private key pairs, $\{k_p, k_s\}$, for authentication. For each traded commodity, e.g., milk and strawberry, a $\mathcal{C}_{com}$ contract is deployed by both $a_m$ and $p_n$ requiring their signatures (\textit{multisig} smart contract). The participants may also deploy off-chain storage in which supplementary data is stored, cf. Section~\ref{sub:layer-view}.

In~$\mathbb{SC}$, we define six supply chain transactions, which provide a link between data layer $\mathbb{L}_{data}$ and blockchain layer $\mathbb{L}_{bc}$. To illustrate how supply chain operations and transactions are used in practice, we select yogurt supply chain as a use case, which is illustrated in Fig.~\ref{fig:sequence-diagram}.

The supply chain process is commenced by importing raw materials into blockchain $\mathbb{B}$ as digital assets. As a complement to $\mathsf{create}$ operation (cf.~\eqref{eq:data-create}), \textbf{\textit{create}} transaction invoked by a producer $p_n$ is defined as follows:
\begin{equation}
\label{eq:tx_create}
    Tx^n_c = [\, Tx^{n}_{ID} \parallel \left<\mathcal{D}^{n,q}_{\mathit{prop}}\right> \parallel \mathit{timestamp} \parallel Sig_{p_n} \,]
\end{equation}
where $Tx^{n}_{ID}$ is the batch ID and $Sig_{p_n}$ is the signature on $hash(\left<\mathcal{D}^{n,q}_{\mathit{prop}}\right> {\parallel} \mathit{timestamp})$ using the signing key $k_{s_p}$ for authentication. $\mathcal{C}_{trm}$ contract assigns a default trust score for $\widehat{t}_{def} = 0$ to $d_{n,q}$. Here, some properties in $\left<\mathcal{D}^{n,q}_{\mathit{prop}}\right>$ may include pointers and the hash of off-chain auxiliary data.

Subsequently, $p_n$ is required to monitor the quality of $d_{n,q}$. Gateway node $\textsl{g}_{n,k}$ submits observational data $\left<\mathbf{v}^{n}_o,\ell_k\right>$ to a blockchain peer node over a secure channel to invoke \textbf{\textit{monitor}} transaction, which is defined as:
\begin{equation}
    Tx^{n,k}_m = [\, Tx^{n,k}_{ID} \parallel \mathbf{v}^{n}_o \parallel \ell_k \parallel \mathit{timestamp} \parallel Sig_{p_n} \,].
\end{equation}
In this transaction, $\mathcal{C}_{trm}$ contract updates the asset's trust score $\widehat{t}_{n,q}$ in each $Tx^{n,k}_m$ transaction, cf.~\eqref{eq:trust-for-assets}. $\mathcal{C}_{trm}$ contract also checks if any temperature reading is outside the expected range, i.e., $\mathcal{V}_{j,i} \notin [\mathrm{T}_{min}, \mathrm{T}_{max}]$ and emits an alert to warn the participant $p_n$ when this occurs.

Suppose producer $p_n$ would like to produce a new batch of strawberry yogurt from a batch of raw milk $d_{n,q}$ and strawberry $d_{n,(q+1)}$, cf.~\eqref{eq:data-produce}. Producer $p_n$ is required to invoke \textbf{\textit{produce}} for capturing the production of strawberry yogurt, which is defined as follows:
\begin{equation}
\label{eq:tx_produce}
    Tx^n_p = [\, Tx^{n}_{ID} \parallel \left<\mathcal{D}^{n,q}_{prop}\right> \parallel \mathit{timestamp} \parallel Sig_{p_n} \,].
\end{equation}
Note that \textit{produce} transaction records the commodity provenance by appending the batch ID's of source commodities to $\left<\mathcal{D}^{n,q}_{prop}\right>$, as well as the source commodities' trust scores, i.e., $\{\widehat{t}_{n,1}, \widehat{t}_{n,2}, \ldots\}$. The new commodity is assigned a new trust score from the average of the source commodities' trust scores.

As a complement to $\mathsf{inspect}$ operation, cf.~\eqref{eq:data-inspect}, food authority $a_m$ invokes \textbf{\textit{inspect}} transaction to permanently store the inspection report, which is defined as:
\begin{equation}
\label{eq:tx_inspect}
    Tx^m_i = [\, Tx^{i,m}_{ID} \parallel e_{m} \parallel H(\mathbf{R}^q) \parallel \mathit{\mathit{timestamp}} \parallel Sig_{p_m} \,]
\end{equation}
where $Tx^{i,m}_{ID}$ is a unique inspection ID and $H(\mathbf{R}^q)$ is the hash of inspection report which is stored off-chain. In \textit{inspect} transaction, $\mathcal{C}_{trm}$ updates the reputation score $\widehat{R}_n$ based on the new endorsement rating scores $e_{m,q}$, cf.~\eqref{eq:participant-reputation}.

\textbf{\textit{Trade}} transaction is a \textit{multisig} transaction, which has to be signed by both seller and buyer to confirm a purchase two parties, cf.~\eqref{eq:data-trade}. The \textit{trade} transaction between participant $p_{n}$ (seller) and $p_{n+1}$ (buyer) is defined as follows:
\begin{equation}
\label{eq:tx_trade}
    Tx_t = [\, Tx^{t}_{ID} \parallel \left<\mathcal{D}^{n,q}_{prop}\right> \parallel \mathbf{tc}_x \parallel \mathit{timestamp} \parallel Sig_p\,]
\end{equation}
where $Sig_p$ corresponds to the signature of both $p_{n}$ and $p_{n+1}$. $\mathcal{C}_{trm}$ validates whether both parties have fulfilled the terms and conditions $\mathbf{tc}_x$, after which $\mathcal{C}_{trm}$ updates both $\widehat{T}_{n}$ and $\widehat{R}_n$ scores accordingly for both parties, cf.~\eqref{eq:participant-trust} and~\eqref{eq:participant-reputation}. Note that supplementary data for \textit{trade} transaction can be stored off-chain, while the hash is stored on-chain.

The final strawberry yogurt product would be imprinted with a QR code that would allow the consumers to query the blockchain $\mathbb{B}$ for the provenance and trust scores for the product. The consumer can access blockchain $\mathbb{B}$ through the REST API in $\mathbb{L}_{app}$ hosted by food authorities. The \textbf{\textit{query}} transaction is defined as: 
\begin{equation}
\label{eq:tx_query}
    Tx_q = [\, \mathit{type} \parallel \mathit{timestamp}\,]
\end{equation}
where $\mathit{type}$ corresponds to trust and provenance of the product or to check the quality of the ingredients. Note that \textit{query} only involves read operation and thus does not alter the state of blockchain $\mathbb{B}$.



\begin{figure*}
    \centering
    \begin{tabularx}{\linewidth}{XXX}
        \includegraphics[width=0.32\textwidth]{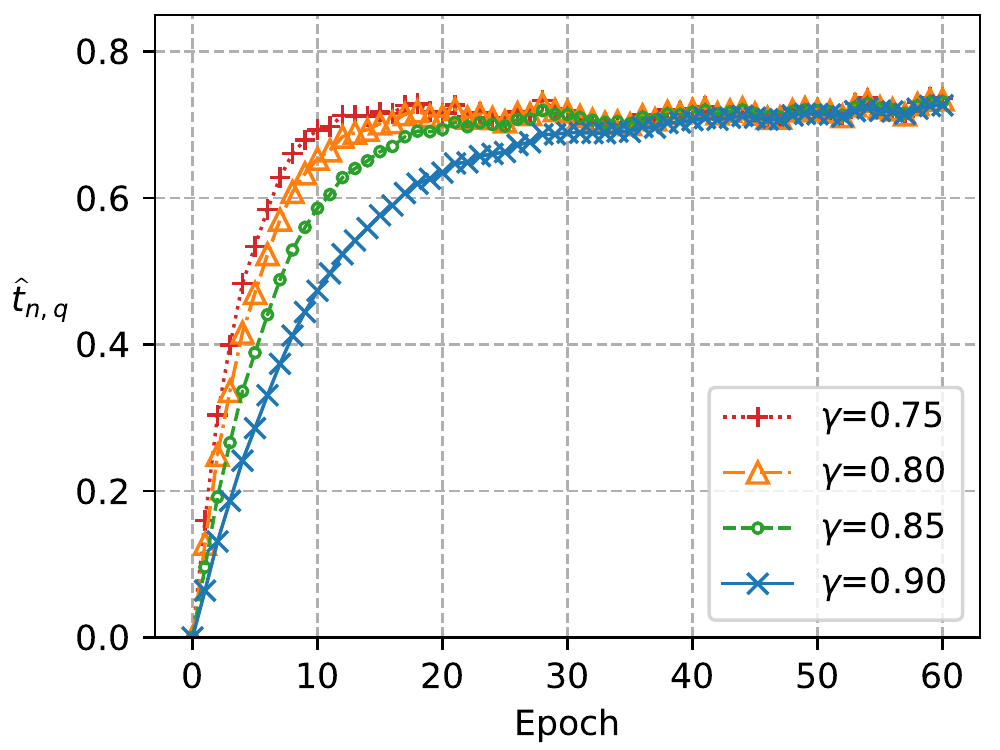}
        \caption{Convergence of $\widehat{t}_{n,q}$ for normal observations with variations of $\gamma$.}
        \label{fig:trust-evolution}
        &
        \includegraphics[width=0.32\textwidth]{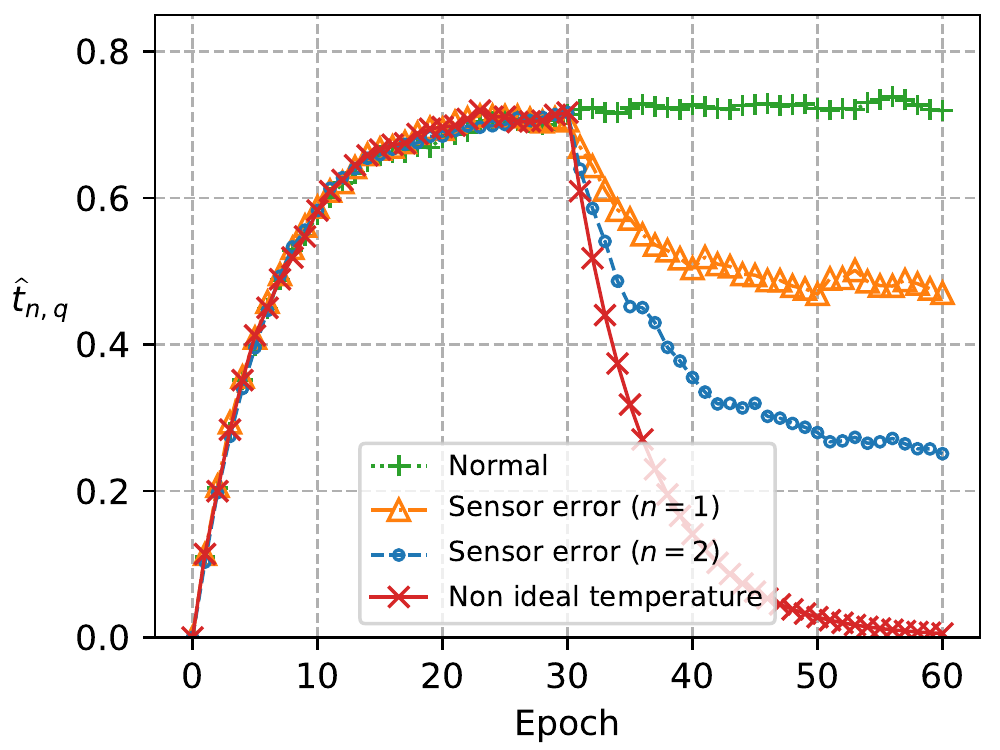}
        \caption{Evolution of $\widehat{t}_{n,q}$ for abnormal storage conditions.}
        \label{fig:faulty-sensor}
        &
        \includegraphics[width=0.32\textwidth]{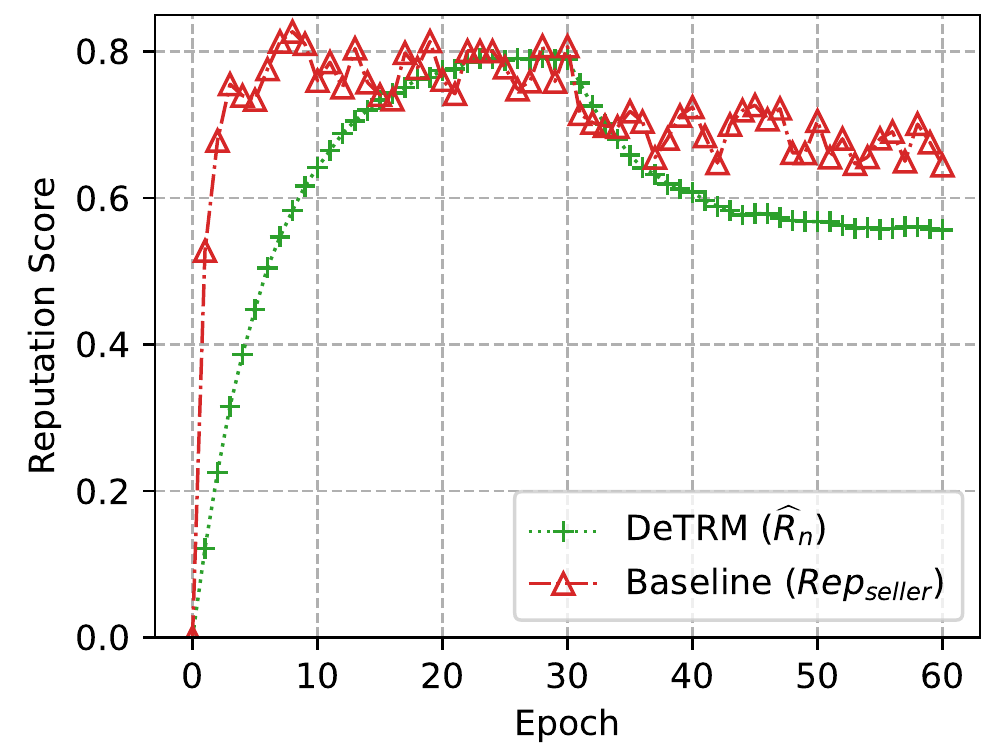}
        \caption{Comparison of reputation score evolution for participant against the baseline~\cite{malik2019}.}
        \label{fig:reputation-comparison}
    \end{tabularx}
\end{figure*}

\begin{figure*}
    \begin{tabularx}{\linewidth}{XX}
        \centering
        \includegraphics[width=0.47\textwidth]{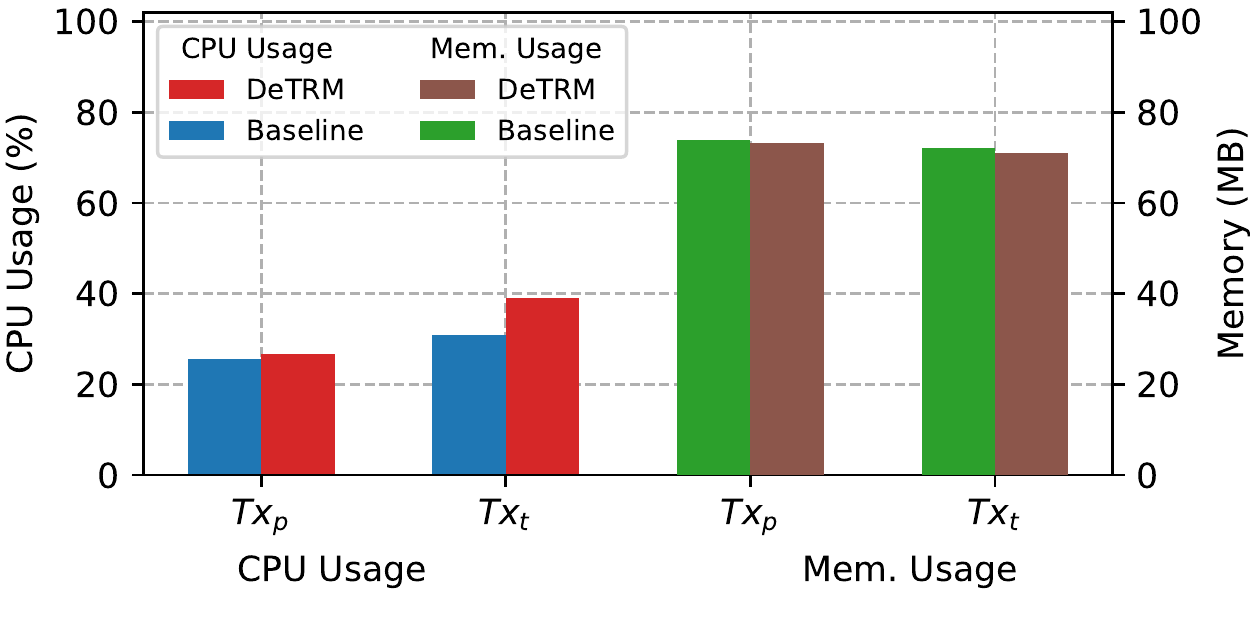}
        \caption{Comparison of CPU and memory consumption for produce and trade transactions between DeTRM and the baseline.}
        \label{fig:cpu-mem-usage}
        &
        \centering
        \includegraphics[width=0.47\textwidth]{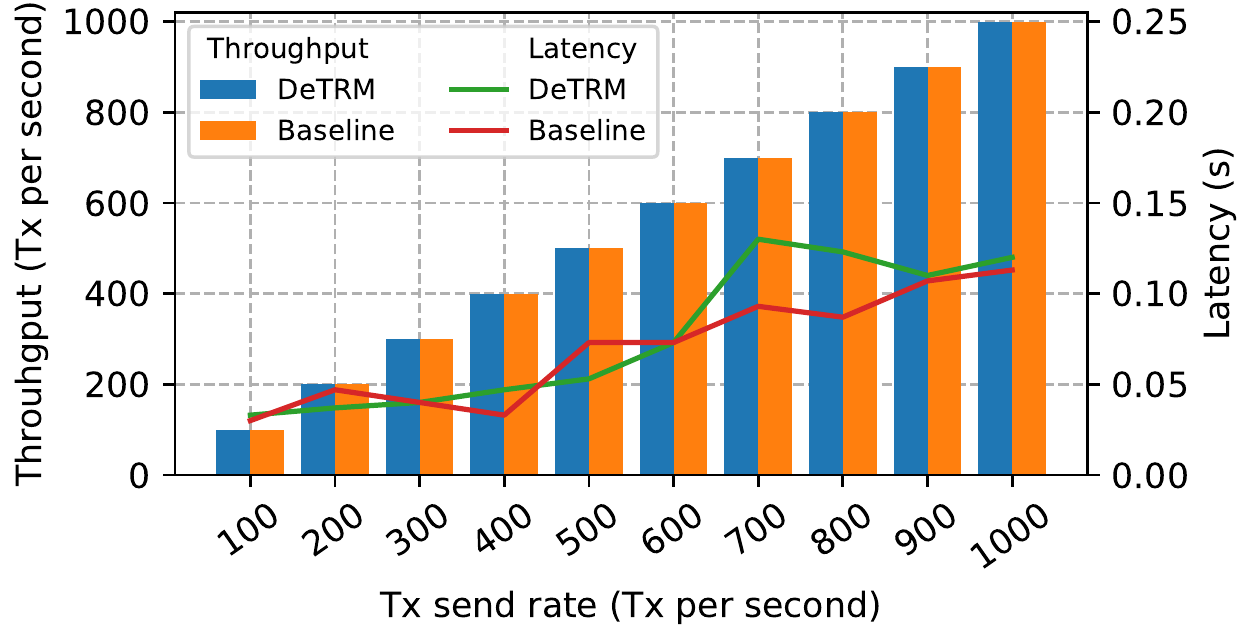}
        \caption{Throughput and latency for trade transactions with varying number of transaction send rate.}
        \label{fig:tx-throughput}
    \end{tabularx}
\end{figure*}

\section{Proof of Concept Evaluation}
\label{sec:performance-evaluation}
In this section, we present proof-of-concept evaluation of DeTRM, which includes evaluation of the proposed TRM and performance evaluation of the blockchain implementation.

\subsection{Implementation Details}
\label{sub:implementation-details}
We developed a proof-of-concept implementation of our proposed framework, DeTRM, on Hyperledger Fabric v2.1, as it natively supports creation of a permissioned blockchain environment and it has built-in compatibility for smart contracts, referred to as chaincodes. We built a private docker container network\footnote{\url{https://www.docker.com/products/docker-desktop}} on an Intel Core i7-9700K CPU 3.6-GHz 8 cores processor with 8 GB memory, running Ubuntu 18.04 LTS operating system, where we configure four distinct organisations to represent a producer, distributor, retailer, and food authority. Here, each organisation runs two blockchain peer nodes to communicate with each other and maintain the blockchain. We consider a fault tolerant consensus mechanism of RAFT protocol~\cite{ongaro2014search} as the ordering service, with with one main communicating channel. We wrote all supply chain operations and transactions as chaincodes in Go programming language\footnote{\url{https://go.dev/}}.
We used Hyperledger Caliper\footnote{\url{https://hyperledger.github.io/caliper/}}, a blockchain benchmark tool, to measure the performance of the blockchain network, e.g., throughput and latency for each transaction.

\subsection{Experimental Results}
\label{sub:experimental-results}
To study the incurred overheads of DeTRM, we compare the evaluation results against a baseline blockchain-based system, which only records provenance and ownership change of commodities without incorporating any TRM operation. In addition, we compare DeTRM against TrustChain to observe how trust and reputation evolves over time~\cite{malik2019}.
    
    \subsubsection{Trust model evaluation}
    We incorporate a decaying constant $\gamma$ in almost all trust and reputation calculation of our trust model, cf.~\eqref{eq:trust-for-assets},~\eqref{eq:participant-trust}, and~\eqref{eq:endorsement-formula}. To study the effect of having different values for $\gamma$, we simulate a scenario where an asset $d_{n,q}$ is properly stored as per the quality and safety standard and plot the evolution of $\widehat{t}_{n,q}$ for $\gamma=\{0.75, 0.80, 0.85, 0.90\}$ in Fig.~\ref{fig:trust-evolution}. While different values of $\gamma$ result in different weightings, $\widehat{t}_{n,q}$ converges to the same upper bound. However, the higher the value of $\gamma$ the more gradual $\widehat{t}_{n,q}$ grows, which indicates that the growth of $\widehat{t}_{n,q}$ can be tuned according to the nature of the commodity. For example, we can set a low value for $\gamma$ to get faster $\widehat{t}_{n,q}$ convergence rate for fast-moving consumer goods, e.g. non-durable food products.
    
    Next, we investigate the evolution of $\widehat{t}_{n,q}$ for four assets under different practical scenarios, e.g., $d_{n,1}$ stored in a normal condition, $d_{n,2}$ has $1$ out of $7$ faulty sensor nodes, $d_{n,3}$ has $2$ out of $7$ faulty sensor nodes and $d_{n,4}$ stored in an environment that does not meet the temperature guidelines. Initially, all assets receive appropriate treatments up to $epoch=30$. We plot the evolution of $\widehat{t}_{n,q}$ in Fig.~\ref{fig:faulty-sensor}, in which we set $\delta_{max} = 1$, $\delta_{min} = 0$, and $\gamma = 0.85$. All scores commence at $\widehat{t}_{n,q}=0$ and continue to increase at an identical rate up to $\mathit{epoch}=30$. While $\widehat{t}_{n,1}$ continues to plateau at the upper bound, $\widehat{t}_{n,2}$, $\widehat{t}_{n,3}$ and $\widehat{t}_{n,4}$ begin to decline from $\mathit{epoch}=30$. Here, different scenarios result in different rates of decline, which is a useful indicator of what is causing the decline of $\widehat{t}_{n,q}$. Temperature out of the ideal condition will trigger significant decline ($\widehat{t}_{n,4}$), while sensor failures produce relatively gradual decline ($\widehat{t}_{n,2}$ and $\widehat{t}_{n,3}$). Note that, $\mathcal{C}_{trm}$ contract will also emit warning notifications to alert the owner, e.g., $p_n$, of this abnormality. Subsequently, $p_n$ can take immediate actions to remedy the condition, such as replacing or re-calibrating the sensor nodes.
    
    To see how DeTRM compares with the baseline~\cite{malik2019}, we consider similar scenario as in Fig.~\ref{fig:faulty-sensor}, where an asset is stored in a temperature out of the ideal condition. We use different modelling of the reputation score and plot the result in Fig.~\ref{fig:reputation-comparison}, in which we use the parameters as recommended in~\cite{malik2019}.
    Ideally, the reputation should grow gradually followed by a steep decline if an adversarial encounter happens. Our reputation model ($\widehat{R}_n$) exhibits stable and gradual increase of reputation score, reaching the upper limit near $\mathit{epoch}=30$. Our model also shows a relatively higher rate of decline. On the other hand, the reputation model of the baseline ($Rep_{sens}$) abruptly hits the upper limit when $\mathit{epoch}<10$ and seems to be unstable. In addition, the decline during non ideal temperature is hardly noticeable due to undesirable fluctuations. In summary, we argue that our model exhibits better trust score evolution to model the reputation for a supply chain participant.

    \subsubsection{Performance of blockchain implementation}
    In DeTRM, we consider \textit{produce} and \textit{trade} transactions, i.e., $Tx_p$ and $Tx_t$, as important supply chain events. In these transactions, we incorporate the TRM in addition to recording of ownership change, which is a key distinction from other blockchain-based SCMS. To study the associated overheads of implementing additional TRM processes in an SCMS, we compare the resource utilisation for $Tx_p$ and $Tx_t$. We utilised Caliper monitor module to measure CPU and memory consumption, where we supplied a load of $500$ transactions per second (tx/second) to the blockchain network. Specifically, we looked at CPU and memory consumption and compared the results in Fig.~\ref{fig:cpu-mem-usage}. DeTRM introduced insignificant overheads in CPU usage, which are relatively negligible for $Tx_p$. Furthermore, DeTRM incurs a slightly lower memory footprint than the baseline.  

    Next, we evaluate the performance of DeTRM against the baseline for executing trade transactions $Tx_t$ where we increase the transaction load from $100$ to $1000$ tx/second. We plot the results in Fig.~\ref{fig:tx-throughput}.
    We measure throughput as the rate at which all requested transactions are executed successfully, while we consider latency as the overall time for completing a transaction. Note that this definition excludes network latency, which is generally influenced by various external factors. For all values of transaction send rate, DeTRM achieved similar throughput as the baseline, despite some additional TRM processes. In addition, DeTRM demonstrated insignificant latency overheads, which are hardly noticeable especially for lower transaction send rate. We note that these performance measures are achieved with our hardware specifications, cf.~Section~\ref{sub:implementation-details}, and would vary depending on the exact hardware specifications and configurations.

\section{Conclusion}
\label{sec:conclusion}
In this paper, we presented DeTRM, a decentralised TRM framework for blockchain-based SCMS, which aims to overcome trust issues in data and behaviour of supply chain participants.
We designed a layered architecture, namely physical, data, blockchain and application, where trust is sourced from sensor observations in the physical layer and adherence to trading agreements in blockchain layer.
Experimental results indicated that DeTRM is feasible and only incurs minimal overheads compared to the baseline.
For future work, we aim to conduct more extensive evaluation which considers a real world SCMS and investigate the actual performance. In addition, we plan to introduce fees and incentives mechanisms in the SCMS and study the economic model and game theory behind these mechanisms.

\section*{Acknowledgements}
\noindent The authors acknowledge the support of the Commonwealth of Australia and Cyber Security Cooperative Research Centre for this work. This work was also supported by the Institute of Information \& Communications Technology Planning \& Evaluation (IITP) grant funded by Korea (MSIT) (2020-0-01594, PSAI industry-academic joint research and education program) and the ITRC (Information Technology Research Center) support program (IITP-2021-2017-0-01633).

\bibliographystyle{IEEEtran}
\bibliography{./mybib}

\begin{thebibliography}{10}
\providecommand{\url}[1]{#1}
\csname url@samestyle\endcsname
\providecommand{\newblock}{\relax}
\providecommand{\bibinfo}[2]{#2}
\providecommand{\BIBentrySTDinterwordspacing}{\spaceskip=0pt\relax}
\providecommand{\BIBentryALTinterwordstretchfactor}{4}
\providecommand{\BIBentryALTinterwordspacing}{\spaceskip=\fontdimen2\font plus
\BIBentryALTinterwordstretchfactor\fontdimen3\font minus
  \fontdimen4\font\relax}
\providecommand{\BIBforeignlanguage}[2]{{%
\expandafter\ifx\csname l@#1\endcsname\relax
\typeout{** WARNING: IEEEtran.bst: No hyphenation pattern has been}%
\typeout{** loaded for the language `#1'. Using the pattern for}%
\typeout{** the default language instead.}%
\else
\language=\csname l@#1\endcsname
\fi
#2}}
\providecommand{\BIBdecl}{\relax}
\BIBdecl

\bibitem{juma2019}
H.~Juma, K.~Shaalan, and I.~Kamel, ``A {{Survey}} on {{Using Blockchain}} in
  {{Trade Supply Chain Solutions}},'' \emph{IEEE Access}, vol.~7, pp.
  184\,115--184\,132, 2019.

\bibitem{gonczol2020}
P.~Gonczol, P.~Katsikouli, L.~Herskind, and N.~Dragoni, ``Blockchain
  {{Implementations}} and {{Use Cases}} for {{Supply Chains-A Survey}},''
  \emph{IEEE Access}, vol.~8, pp. 11\,856--11\,871, 2020.

\bibitem{rejeb2019}
A.~Rejeb, J.~G. Keogh, and H.~Treiblmaier, ``Leveraging the {{Internet}} of
  {{Things}} and {{Blockchain Technology}} in {{Supply Chain Management}},''
  \emph{Future Internet}, vol.~11, no.~7, p. 161, Jul. 2019.

\bibitem{tajima2007}
M.~Tajima, ``Strategic value of {{RFID}} in supply chain management,''
  \emph{Journal of Purchasing and Supply Management}, vol.~13, no.~4, pp.
  261--273, Dec. 2007.

\bibitem{sahai2020}
S.~Sahai, N.~Singh, and P.~Dayama, ``Enabling {{Privacy}} and {{Traceability}}
  in {{Supply Chains}} using {{Blockchain}} and {{Zero Knowledge Proofs}},'' in
  \emph{2020 {{IEEE International Conference}} on {{Blockchain}}
  ({{Blockchain}})}.\hskip 1em plus 0.5em minus 0.4em\relax {Rhodes Island,
  Greece}: {IEEE}, Nov. 2020, pp. 134--143.

\bibitem{malik2019}
S.~Malik, V.~Dedeoglu, S.~S. Kanhere, and R.~Jurdak, ``{{TrustChain}}: Trust
  {{Management}} in {{Blockchain}} and {{IoT Supported Supply Chains}},'' in
  \emph{2019 {{IEEE International Conference}} on {{Blockchain}}
  ({{Blockchain}})}, Jul. 2019, pp. 184--193.

\bibitem{malik2018}
S.~Malik, S.~S. Kanhere, and R.~Jurdak, ``{{ProductChain}}: Scalable
  {{Blockchain Framework}} to {{Support Provenance}} in {{Supply Chains}},'' in
  \emph{2018 {{IEEE}} 17th {{International Symposium}} on {{Network Computing}}
  and {{Applications}} ({{NCA}})}.\hskip 1em plus 0.5em minus 0.4em\relax
  {Cambridge, MA}: {IEEE}, Nov. 2018, pp. 1--10.

\bibitem{powell2021a}
W.~Powell, M.~Foth, S.~Cao, and V.~Natanelov, ``Garbage in garbage out: The
  precarious link between {{IoT}} and blockchain in food supply chains,''
  \emph{Journal of Industrial Information Integration}, p. 100261, Aug. 2021.

\bibitem{shahid2020}
A.~Shahid, A.~Almogren, N.~Javaid, F.~A. {Al-Zahrani}, M.~Zuair, and M.~Alam,
  ``Blockchain-{{Based Agri}}-{{Food Supply Chain}}: A {{Complete Solution}},''
  \emph{IEEE Access}, vol.~8, pp. 69\,230--69\,243, 2020.

\bibitem{li2020a}
H.~Li, K.~Gai, L.~Zhu, P.~Jiang, and M.~Qiu, ``Reputation-{{Based Trustworthy
  Supply Chain Management Using Smart Contract}},'' in \emph{Algorithms and
  {{Architectures}} for {{Parallel Processing}}}, ser. Lecture {{Notes}} in
  {{Computer Science}}, M.~Qiu, Ed.\hskip 1em plus 0.5em minus 0.4em\relax
  {Cham}: {Springer International Publishing}, 2020, pp. 35--49.

\bibitem{zhang2021}
D.~Zhang, X.~Xu, L.~Zhu, and H.-Y. Paik, ``A {{Process Adaptation Framework}}
  for {{Blockchain}}-{{Based Supply Chain Management}},'' in \emph{2021 {{IEEE
  International Conference}} on {{Blockchain}} and {{Cryptocurrency}}
  ({{ICBC}})}.\hskip 1em plus 0.5em minus 0.4em\relax {Sydney, Australia}:
  {IEEE}, May 2021, pp. 1--9.

\bibitem{putra2021a}
G.~D. Putra, V.~Dedeoglu, S.~S. Kanhere, and R.~Jurdak, ``Blockchain for
  {{Trust}} and {{Reputation Management}} in {{Cyber}}-physical {{Systems}},''
  \emph{arXiv:2109.07721 [cs]}, Sep. 2021.

\bibitem{putra2021b}
G.~D. Putra, V.~Dedeoglu, S.~S. Kanhere, R.~Jurdak, and A.~Ignjatovic,
  ``Trust-{{Based Blockchain Authorization}} for {{IoT}},'' \emph{IEEE
  Transactions on Network and Service Management}, vol.~18, no.~2, pp.
  1646--1658, Jun. 2021.

\bibitem{camilo2020}
G.~F. Camilo, G.~A.~F. Rebello, L.~A.~C. {de Souza}, and O.~C. M.~B. Duarte,
  ``A {{Secure Personal}}-{{Data Trading System Based}} on {{Blockchain}},
  {{Trust}}, and {{Reputation}},'' in \emph{2020 {{IEEE International
  Conference}} on {{Blockchain}} ({{Blockchain}})}, Nov. 2020, pp. 379--384.

\bibitem{moinet2017}
A.~Moinet, B.~Darties, and J.-L. Baril, ``Blockchain based trust \&
  authentication for decentralized sensor networks,'' \emph{arXiv:1706.01730
  [cs]}, Jun. 2017.

\bibitem{mortimore2013haccp}
S.~Mortimore and C.~Wallace, \emph{HACCP: A practical approach}.\hskip 1em plus
  0.5em minus 0.4em\relax Springer Science \& Business Media, 2013.

\bibitem{frolik2001}
J.~Frolik, M.~Abdelrahman, and P.~Kandasamy, ``A confidence-based approach to
  the self-validation, fusion and reconstruction of quasi-redundant sensor
  data,'' \emph{IEEE Transactions on Instrumentation and Measurement}, vol.~50,
  no.~6, pp. 1761--1769, Dec. 2001.

\bibitem{androulaki2018}
E.~Androulaki, A.~Barger, V.~Bortnikov, C.~Cachin, K.~Christidis, A.~De~Caro,
  D.~Enyeart, C.~Ferris, G.~Laventman, Y.~Manevich, S.~Muralidharan, C.~Murthy,
  B.~Nguyen, M.~Sethi, G.~Singh, K.~Smith, A.~Sorniotti, C.~Stathakopoulou,
  M.~Vukoli{\'c}, S.~W. Cocco, and J.~Yellick, ``Hyperledger fabric: A
  distributed operating system for permissioned blockchains,'' in
  \emph{Proceedings of the {{Thirteenth EuroSys Conference}}}, ser. {{EuroSys}}
  '18.\hskip 1em plus 0.5em minus 0.4em\relax {New York, NY, USA}: {Association
  for Computing Machinery}, Apr. 2018, pp. 1--15.

\bibitem{brown2016corda}
R.~G. Brown, J.~Carlyle, I.~Grigg, and M.~Hearn, ``Corda: an introduction,''
  \emph{R3 CEV, August}, vol.~1, p.~15, 2016.

\bibitem{8543246}
M.~A. {Ferrag}, M.~{Derdour}, M.~{Mukherjee}, A.~{Derhab}, L.~{Maglaras}, and
  H.~{Janicke}, ``Blockchain technologies for the internet of things: Research
  issues and challenges,'' \emph{IEEE IoT Journal}, vol.~6, no.~2, pp.
  2188--2204, 2019.

\bibitem{dedeoglu2019}
V.~Dedeoglu, R.~Jurdak, G.~D. Putra, A.~Dorri, and S.~S. Kanhere, ``A trust
  architecture for blockchain in {{IoT}},'' in \emph{Proceedings of the 16th
  {{EAI International Conference}} on {{Mobile}} and {{Ubiquitous Systems}}:
  Computing, {{Networking}} and {{Services}}}.\hskip 1em plus 0.5em minus
  0.4em\relax {Houston Texas USA}: {ACM}, Nov. 2019, pp. 190--199.

\bibitem{dierks2008transport}
T.~Dierks and E.~Rescorla, ``The transport layer security (tls) protocol
  version 1.2,'' 2008.

\bibitem{ongaro2014search}
D.~Ongaro and J.~Ousterhout, ``In search of an understandable consensus
  algorithm,'' in \emph{2014 $\mathit{USENIX}$ Annual Technical Conference
  ($\mathit{USENIX\,ATC}$ 14)}, 2014, pp. 305--319.

\end{thebibliography}

\end{document}